\documentclass[draft]{agujournal2019}
\usepackage{url} 
\usepackage{lineno}
\usepackage[inline]{trackchanges} 
\usepackage{soul}
\usepackage{siunitx}
\DeclareSIUnit\year{year}
\DeclareSIUnit\years{years}
\DeclareSIUnit\ppm{ppm}
\draftfalse

\journalname{JGR: Machine Learning and Computation}
\begin{document}

\title{Applying the ACE2 Emulator to SST Green's Functions for the E3SMv3 Global Atmosphere Model}
\authors{Elynn Wu\affil{1}, Finn Rebassoo\affil{2}, Pappu Paul\affil{3}, Cristian Proistosescu\affil{3,4}, Jacqueline Nugent\affil{5}, 
Daniel McCoy\affil{5}, Peter Caldwell\affil{2}, Christopher S. Bretherton\affil{1}}

\affiliation{1}{Allen Institute for Artificial Intelligence (Ai2), Seattle, WA, USA}
\affiliation{2}{Lawrence Livermore National Laboratory, Livermore, CA USA}
\affiliation{3}{Department of Climate, Meteorology, and Atmospheric Sciences, University of Illinois at Urbana‐Champaign, Champaign, IL, USA}
\affiliation{4}{Department of Earth Sciences and Environmental Change, University of Illinois at Urbana‐Champaign, Champaign, IL, USA}
\affiliation{5}{Department of Atmospheric Science, University of Wyoming, Laramie, WY, USA}
\correspondingauthor{Elynn Wu}{elynnw@allenai.org}
\begin{keypoints}
\item The Ai2 Climate Emulator accurately emulates top-of-atmosphere (TOA) radiative response to local sea surface temperature (SST) anomalies
\item ACE and EAMv3's global TOA radiation sensitivity to all SST patch anomalies are qualitatively similar but differ in spatial details
\item Historical reconstruction of TOA radiation from SST anomalies are captured by both ACE and EAMv3
\end{keypoints}
\section*{Abstract}
Green's functions are a useful technique for interpreting atmospheric state responses to changes in the spatial pattern of sea surface temperature (SST). Here we train version 2 of the Ai2 Climate Emulator (ACE2) on reference historical SST simulations of the  US Department of Energy's EAMv3 global atmosphere model.  We compare how well the SST Green's functions generated by ACE2 match those of EAMv3, following the protocol of the Green's Function Model Intercomparison Project (GFMIP). The spatial patterns of top-of-atmosphere (TOA) radiative response from the individual GFMIP SST patch simulations are similar for ACE and the EAMv3 reference.  The derived sensitivity of global net TOA radiation sensitivity to SST patch location is qualitatively similar in ACE as in EAMv3, but there are statistically significant discrepancies for some SST patches, especially over the subtropical northeast Pacific.  These discrepancies may reflect insufficient diversity in the SST patterns sampled over the course of the EAMv3 AMIP simulation used for training ACE. Both ACE and EAMv3 Green's functions reconstruct the historical record of the global annual-mean TOA radiative flux from a reference EAMv3 AMIP simulation reasonably well. Notably, under our configuration and compute resources, ACE achieves these results approximately 100 times faster in wall-clock time compared to EAMv3, highlighting its potential as a powerful and efficient tool for tackling other computationally intensive problems in climate science.

\section*{Plain Language Summary}
The Green's Function Model Intercomparison Project (GFMIP) is a standardized framework used to study the atmospheric response to changes in sea surface temperature. Traditionally, these experiments are conducted using physics based climate models, which are computationally expensive to run. In this study, we use a machine-learning based emulator-- the Ai2 climate emulator version 2 (ACE2)-- to carry out the GFMIP protocol. ACE2 completes the same experiment roughly 100 times faster than the physics based climate model and produces qualitatively similar top-of-atmosphere radiative response. While some differences remain between ACE2 and the physics-based model, these are likely due to insufficient training data. We believe that ACE2's remaining biases can be overcome in the near future, making it an efficient tool for addressing other computationally intensive problems in climate science.

\section{Introduction}
Green’s functions have proven a useful tool for interpreting atmospheric state responses to changes in the spatial pattern of sea surface temperature (SST). This technique, introduced by \citeA{branstator_analysis_1985} and \citeA{barsugli_global_2002}, has been adopted over the last decade by many climate modeling centers \cite{zhou_analyzing_2017, dong_attributing_2019, zhang_using_2023, alessi_surface_2023} to provide insights into how spatial patterns of warming in SST affect the global radiative feedback on greenhouse warming, also called the “pattern effect.” This has emerged as a key issue in relating historical observations of global warming and cloud trends to future climate projections.

Recently, \citeA{bloch-johnson_greens_2024}, hereafter BJ24, outlined the Green's Function Model Intercomparison Project (GFMIP), a protocol to standardize the application of Green’s functions so as to identify true differences among responses of various climate models to patterned SST anomalies that are not due to inconsistencies in experimental setup. In this protocol, 218 10-year patch simulations, using 109 warmed-SST and 109 cooled-SST patterns to check for response linearity, plus a 20-year control simulation, are requested, totaling 2200 simulation years.  For a modern full-physics climate model with a typical horizontal grid resolution of 100 km, this is a substantial computational expense. For EAMv3 (the atmospheric component of E3SMv3, the recently-released version 3 of the Energy Exascale Earth System Model developed by the U.S. Department of Energy) \cite{xie2025energy}, running the Green's Function simulations required 8.15 million core hours, using 8 Nodes on the Derecho computing system at the National Center for Atmospheric Research (NCAR)-Wyoming Supercomputing Center (NWSC). Each patch simulation was run on 8 CPU nodes (1024 processors), with a throughput of 7 simulated years per day. The simulation lengths in the GFMIP protocol are thus a compromise between affordability and accuracy of radiative response.  Longer patch and control simulations would reduce uncertainties in the climate model response due to internal climate variability but would further raise the computational burden. 

BJ24 found that the GFMIP protocol adequately predicts the response of a global climate model’s net global-mean top-of-atmosphere (TOA) radiative flux response to SST perturbations seen in the historical record.
The GFMIP protocol also shows that different climate models show qualitatively similar sensitivities of global radiative response to patch location, but with substantial quantitative differences presumably due mainly to cloud-related parameterizations.  

While traditional climate models offer physical insight into the pattern effect, machine learning has the potential to greatly accelerate the simulations needed for GFMIP. The Ai2 Climate Emulator (ACE) \cite{WattMeyer2023, watt-meyer_ace2_2024}, an atmospheric model emulator based on machine learning, can carry out the GFMIP simulation suite in 2.3 wall clock days using one NVIDIA A100 GPU. This is less than 1\% of the 331 total wall-clock days to run the full suite of GFMIP simulations with EAMv3 on 8 Derecho nodes. 
Training ACE on a reanalysis or on a 40-year AMIP simulation of DOE's EAMv3 (as detailed below) is similarly quick, albeit with 16 A100 GPUs. 

Thus, it is natural to ask whether a machine learning based climate emulator is ready for this task. Recently, \citeA{loon_reanalysis-based_2025} used three previously published ACE models with an approximately 1$^\circ$ latitude/longitude grid to perform GFMIP simulations. Two were trained to emulate output from global atmosphere models developed at major climate modeling centers, forced with a repeating annual cycle of SST \cite{WattMeyer2023, Duncan2024}.  The third, ACE2-ERA5, was a more recent version of ACE trained on ERA5 reanalysis of historical climate from 1940-2020 \cite{watt-meyer_ace2_2024}. None of these ACE models were trained on any GFMIP-like SST patch simulations. \citeA{loon_reanalysis-based_2025} found that ACE2-ERA5 produced a qualitatively reasonable physical sensitivity map of TOA atmospheric radiative response but likely underestimated the radiative response to historical warming. This is likely due to a mixture of ACE2-ERA5 not properly learning the sensitivity to SST and struggling to do out-of-sample generalization well. Similar results are found in the other two models, though one of them shows a much noisier sensitivity map. However, they did not have a precise ground truth for their ACE patch simulations, making attribution of discrepancies challenging. 

In this study, we build upon BJ24 and \citeA{loon_reanalysis-based_2025} by executing the GFMIP protocol using a version of ACE2 trained on 1970-2020 historical SST-forced (`AMIP-style') simulations with EAMv3. 
We also run the same set of GFMIP patch and control simulations with EAMv3 in order to compare in detail how well ACE emulates the underlying model. 

\section{Data and Methods}
\subsection{ACE2-EAMv3 Training Overview}
\label{sec:ace-eamv3-training}
Our training data is from an AMIP-style \cite{gates_overview_1999, Eyring2016} simulation with EAMv3 from 1970-2020. It is configured to run with E3SM's F2010 component set, except for using AMIP SSTs. ACE is an autoregressive machine learning climate emulator \cite{WattMeyer2023} with 6-hourly temporal resolution and 1$^\circ$ horizontal resolution. Here, we follow the latest version of ACE2's training protocol described by \citeA{watt-meyer_ace2_2024} and \citeA{clark_ace2-som_2024}. This differs from training in \citeA{Duncan2024} in using EAMv3 instead of EAMv2 as the reference model for generating training data and four ACE2 upgrades to our training protocol:  (1) using 51 years of historical SSTs including multiple ENSO cycles and a global warming trend, rather than a repeating annual cycle of SST, (2) optimizing losses from two 6-hourly time steps ahead, (3) using a larger embedding dimension of 384, and (4) enforcing global conservation of dry air mass and moisture.  Unlike the previous ACE2 studies, both the EAMv3 reference simulation and the emulator are forced with constant 2010 CO$_2$ concentrations. We train ACE for 50 epochs and the training takes 1.7 hours per epoch on four 4$\times$A100 GPU nodes on Perlmutter. 

During training, we run an `inline inference' at the end of each epoch. This inference is performed from 1970 to 2020 with the same data on which we train. The best model checkpoint is chosen at the epoch where the inline inference has the lowest mean root mean square error (RMSE) across all predicted variables. This procedure is described in ``Checkpoint selection based on climate skill" in \citeA{watt-meyer_ace2_2024}. 

We train ACE2 with four random seeds; the epoch selected via inline inference for the best model checkpoint varies between these seeds. The random seeds had similar training and validation losses across the training ensemble, and three out of four seeds show similar metrics for the inline inference. 

To determine the best random seed we run a standalone inference from 1970 to 2020 using the checkpoint determined during training. The best random seed (and the one used for the rest of this paper) is selected as the one with the least time-mean RMSE in the net top of the atmosphere radiation for this standalone inference.  

\begin{figure}[h]
    \centering
    \includegraphics[width=\textwidth]{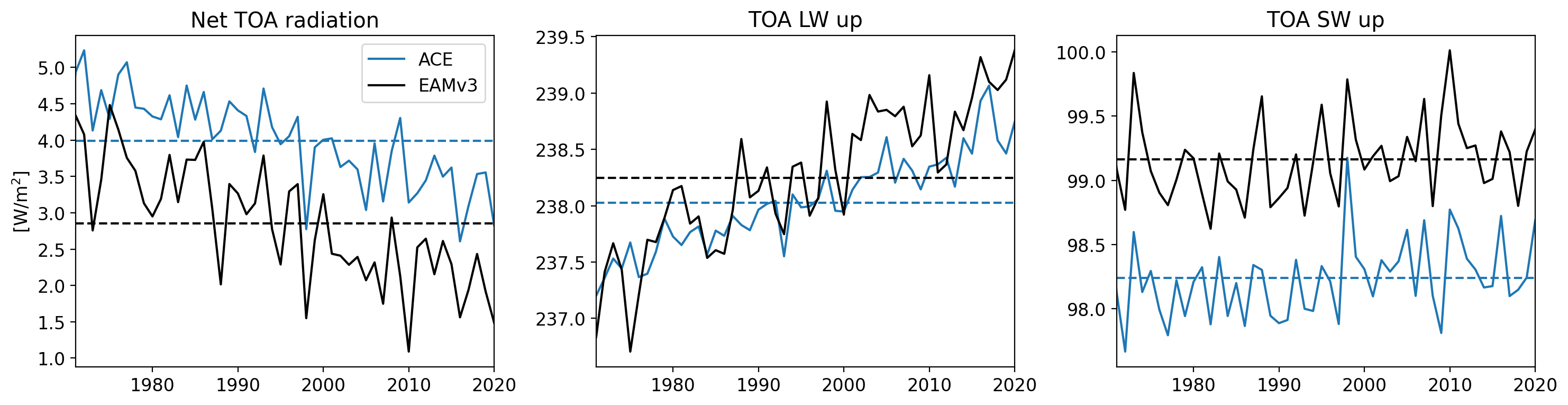}
    \caption{Time series of net TOA (positive downward), TOA upward longwave, and TOA upward shortwave radiation for ACE-EAMv3 and EAMv3 over 51 years of evaluation. The dashed lines show the mean value of each time series over the 51 years. \label{fig:training-rad-time-series}}
\end{figure}

Figure \ref{fig:training-rad-time-series} shows the time series of global-mean net TOA radiation $\overline{N}$ (positive downward; the overline denotes a global average),  TOA upward longwave (LW) radiation, and TOA upward shortwave (SW) radiation for EAMv3 and for an ACE2-EAMv3 simulation with the chosen seed and checkpoint, initialized from the EAMv3 simulation at the start of 1970. ACE2-EAMv3 captures the global trend in the net TOA radiation, with a mean bias of about 1~W/m$^2$ coming mostly from the TOA SW radiation. ACE2-EAMv3's interannual variability in both TOA radiation components (upward longwave and upward shortwave) is highly correlated with the reference model, but with somewhat reduced amplitude, similar to \citeA{watt-meyer_ace2_2024}. 

\begin{figure}[h]
    \centering
    \includegraphics[width=\textwidth]{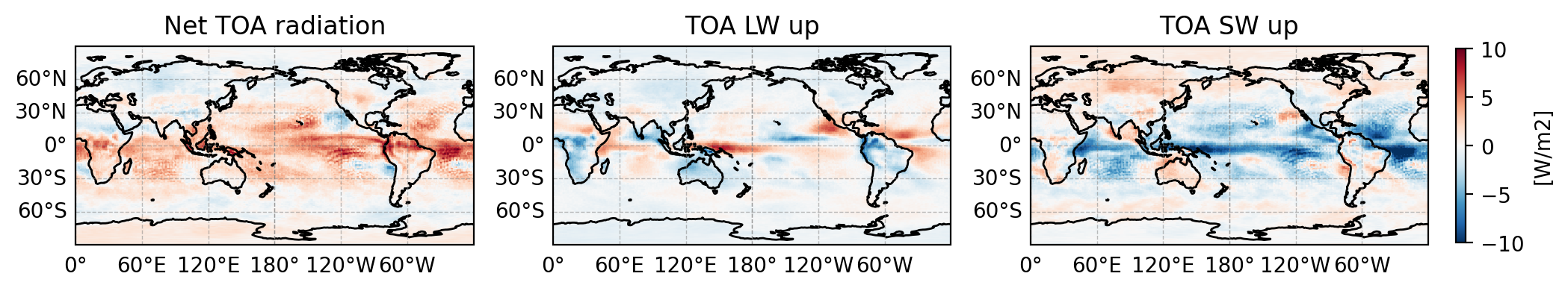}
    \caption{Maps of emulator minus reference bias in net TOA, TOA upward longwave, and TOA upward shortwave radiation averaged over 1970-2020 AMIP simulations.   \label{fig:training-rad-map}}
\end{figure}

Figure \ref{fig:training-rad-map} shows maps of 51-year mean spatially-resolved biases of ACE2-EAMv3 net, LW and SW TOA radiation vs. the EAMv3 reference AMIP simulation.  The net radiation biases are everywhere less than 10 W/m$^2$.  They are largest in low latitudes, where they are systematically positive (a `dim cloud' bias) outside of stratocumulus regions, especially over the tropical eastern Pacific and Atlantic Ocean.

\subsection{GFMIP simulations}
\label{sec:gfmip-simulations}
We follow the GFMIP protocol and notation in Sections 2 and 3 of BJ24. The control simulation of EAMv3 is run for 21 years with the first year as spin-up, followed by all patch simulations conducted for 10 years from the end of the control run in accordance with the GFMIP protocol. Given the high computational cost of running EAMv3, only a subset of simulations are extended up to 40 years for further analysis: the control run and the three patches: tropical ascent (0$^\circ$N, 140$^\circ$E), tropical subsidence (20$^\circ$S, 260$^\circ$E), and extratropical (40$^\circ$N, 180$^\circ$E). The spatial structure of the SST anomaly patches follows their Eqn. (1).  Their patch centers $\phi_0, \theta_0$ are located every $\delta \phi = 10^\circ$ of latitude $\phi$ and every $\delta \theta = 40^\circ$ of longitude $\theta$.
Warm and cold SST patches have central amplitudes $A_p = \pm 2$~K, meridional and zonal widths equal to the spacing between patch centers, and SST perturbations decreasing smoothly and symmetrically to zero at the patch edges:
\begin{equation}
  \Delta SST_p(\phi,\theta) = A_p \cos^2(\pi \min[|\phi - \phi_0|,0.5]/\delta \phi) \cos^2(\pi\min[|\theta - \theta_0|,0.5]/\delta \theta).
  \label{eq:DeltaSSTp}
\end{equation}
Sea ice is specified to follow a fixed climatological seasonal cycle. Following GFMIP specifications, only ice-free ocean grid points are considered in our analysis. We define ocean grid points as ocean fraction greater than 50\% and ice-free as annual maximum sea-ice concentration less than 0.001.

For ACE2-EAMv3, 
we use the same random seed and inference checkpoint as for the AMIP results presented in Section \ref{sec:ace-eamv3-training} to perform control and patch simulations. SST and sea ice fraction are taken directly from GFMIP's website and re-gridded to ACE's 1$^\circ$ Gaussian grid. For TOA incoming solar radiation, we use the same annually repeated value from the F2010 component set as in training. We follow the same patch specification in GFMIP, with a total of 109 patches and using +2~K and -2~K perturbations. Since running ACE is computationally inexpensive, we run our control and all patch simulations to a total of 40 years to better account for interannual variability, totaling 8760 simulation years.  However, unless otherwise stated, ACE2-EAMv3 results are based on the first 10 simulated years for patches and the first 20 simulated years for the control, consistent with the GFMIP specification.  

Since ACE2-EAMv3 is trained on the AMIP-style EAMv3 reference simulation, all patch simulations are out-of-sample tests of the emulator, because they involve SST patterns somewhat different from those seen in training. It is important to recognize that the natural interannual SST variability sampled in the AMIP training data is limited and does not encompass patch SST perturbations like those used in GFMIP. Thus, we anticipate this will be a challenging generalization problem for any AMIP-trained climate emulator. Using a longer and more customized set of EAMv3 reference simulations for training (e.g. the patch simulations themselves) could dramatically improve the emulator skill.  However, it would also lack the observational grounding of AMIP simulations.  It could also potentially require as many years of reference-model simulation for training as would be require to directly derive the Green's functions, defeating the point of using ACE2 in the first place.

\section{Results}
\subsection{Individual patch skill}
We first examine three representative individual patch simulations from ACE2-EAMv3 (hereafter just termed ACE for brevity) and EAMv3, with warm SST patches of maximum amplitude +2~K centered in tropical ascent, tropical subsidence, and extratropical regions. We chose these three patches to match Fig. 2 of BJ24. 
We use the extended (40~yr) EAMv3 simulations for the control SST distribution and for these three SST patch cases to better characterize their time-mean climatologies and interannual variability. 

\begin{figure}[h]
    \centering
    \includegraphics[width=\textwidth]{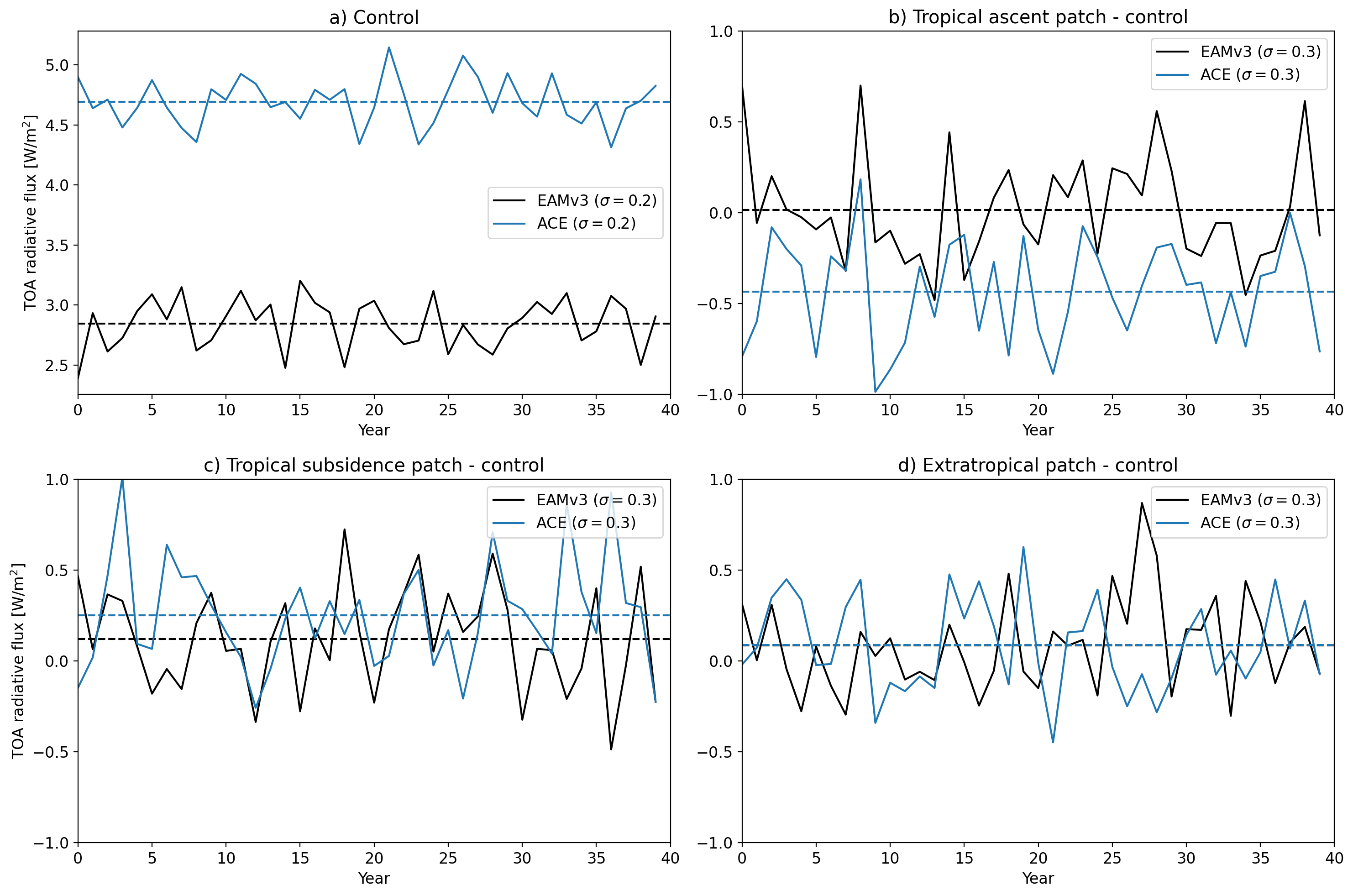}
    \caption{Time series of annual and global mean TOA radiation for the control simulation in ACE and EAMv3, and TOA radiation change from control for tropical ascent, tropical subsidence, and extratropical SST patches. Dash lines indicate the 40~yr average. \label{fig:results-3-patch-rad-time-series}}
\end{figure}

Figure \ref{fig:results-3-patch-rad-time-series} compares the ACE and EAMv3 time series of annual and global-mean TOA net radiation $\overline{N}$ for the control and the three patch simulations.  The dashed lines show the means of these time series.   Figure \ref{fig:results-3-patch-rad-time-series}a shows that the control emulator simulation immediately develops a global 1.8 W/m$^2$ bias vs. EAMv3; this is larger than the time-mean bias of ACE vs. the EAMv3 AMIP reference simulation used for training. However, as seen in Figure \ref{fig:results-3-patch-rad-time-series}b-d, when subtracting ACE and EAMv3's patch simulations from their own controls to get a patch-induced change $\Delta \overline{N}_p$ in net radiation, ACE's time-mean biases vs. EAMv3 are relatively small for all three patches.  Only for the tropical ascent patch does the emulator have a significant time-mean bias in $\Delta \overline{N}_p$ of -0.4~W/m$^2$. 

The year-to-year variability of $\overline{N}_p$ is also reassuringly similar in ACE and EAMv3 for the control simulation, with a standard deviation $\sigma_{\overline{N}} \approx 0.2$~W/m$^2$ and no significant autocorrelation between successive years.  The same is true for the three patch simulations (not shown). Since the three patches are from diverse ocean locations, we regard this interannual variability of $\overline{N}$ as representative of all ocean patch locations.  Its amplitude is consistent with three other climate models mentioned by BJ24, for which the patch and control simulations have interannual standard deviations in the range 0.14--0.24~W/m$^2$.  
For the patch minus control differences shown in Figure \ref{fig:results-3-patch-rad-time-series}b-d, we expect the interannual standard deviation to be $2^{1/2}$ as large, or around 0.3~W/m$^2$, and this is indeed the case for both ACE and EAMv3. 

\begin{figure}[h]
    \centering
    \includegraphics[width=\textwidth]{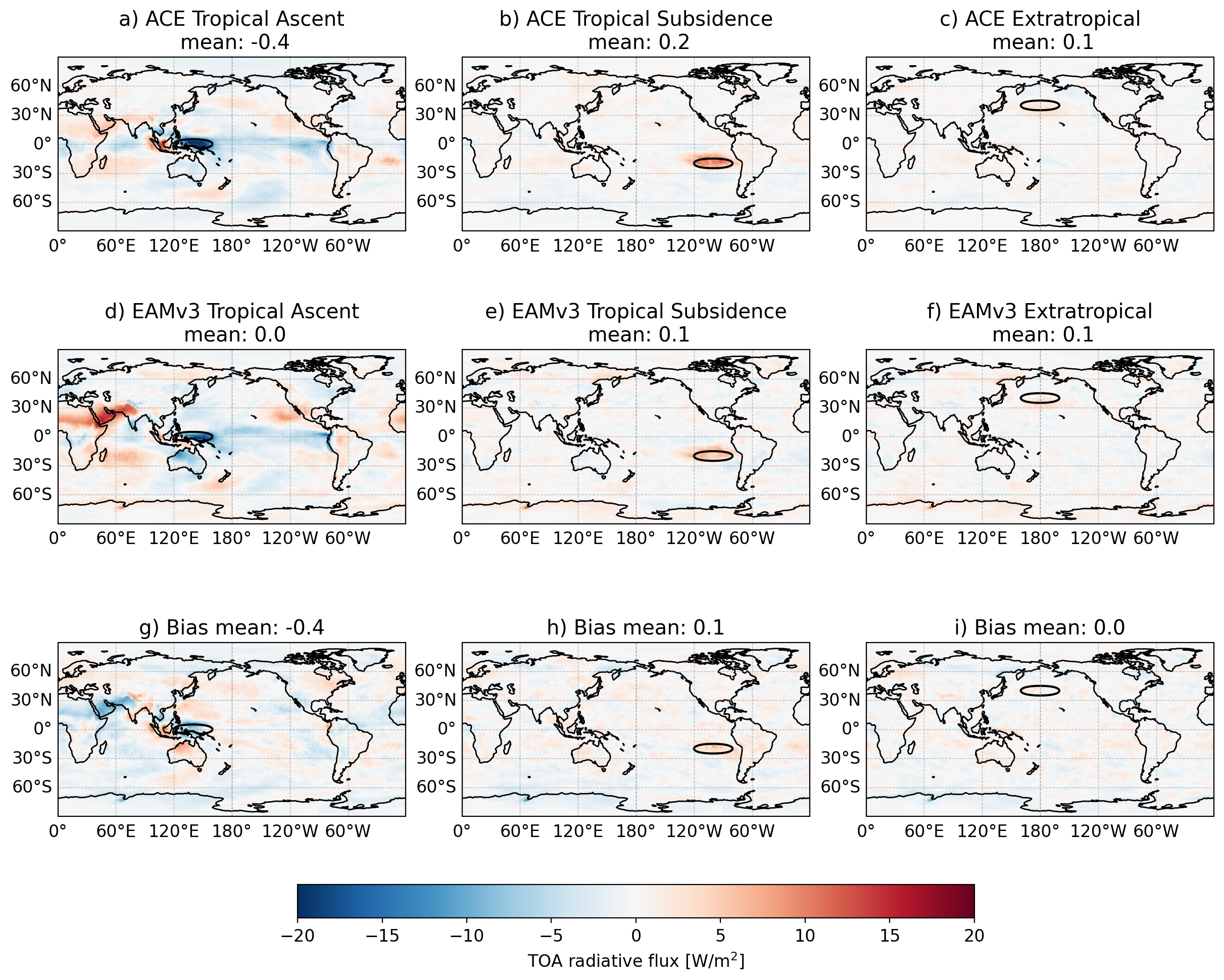}
    \caption{Map of TOA radiative flux changes from control for three patch simulations for ACE (a-c), EAMv3 (d-f), and bias (g-i). Black ellipses indicate half-amplitudes for each of the patch SST perturbation. \label{fig:results-3-patch-rad-map}}
\end{figure}

Figure \ref{fig:results-3-patch-rad-map} shows the 40-year time-mean spatial pattern of patch-induced net TOA radiation change $N$. The ACE and EAMv3 time-mean radiative response to all three SST patches qualitatively agree with the climate model simulation results shown in Figure 2 in BJ24. More important for our purposes, ACE emulates EAMv3 well for all three patch forcings, producing physically plausible results previously explained by \citeA{zhou_analyzing_2017} and others mainly in terms of cloud changes. Remarkably, ACE replicates these radiative responses without ever emulating clouds, purely by learning how their radiative effects correlate with large-scale atmospheric structures that ACE does predict.  

Regionally, the ACE net radiation biases vs. EAMv3 are small (mostly less than 5~W/m$^2$ for the tropical subsidence and extratropical patches), but larger (up to 10~W/m$^2$ over southern Eurasia and the Sahara) for the tropical ascent patch. This emulator accuracy is impressive, 
especially for an out-of-sample test.  However, there are compensating regions of positive and negative net radiation for the EAMv3 reference. 
The magnitude of patch-induced global-time-mean TOA radiation $\overline{N}$ is smaller than 0.5~W/m$^2$ for all three patches, as previously noted.  This is only a few percent of the regional maxima. To capture such a small global-mean effect so as to be reliable for an SST Green's function analysis, ACE2 must emulate EAMv3's time-mean net radiation field very accurately.  We now investigate whether that is the case, and whether this can even be reliably discerned from 10 year patch simulations, given the `noise' of unforced internal variability.

\subsection{Green's function sensitivity maps}

BJ24's Eqns. 2 and 3 use the patch simulation results to assess the sensitivity of TOA net radiation to SST pattern. They use a measure of global radiation sensitivity to ocean-mean patch-induced SST changes:
\begin{equation}
  (d\overline{N}/dSST)_p = \Delta \overline{N}_p/\langle \Delta SST_p \rangle, 
  \label{eq:dRdSST}
\end{equation}
where $\langle \Delta SST_p \rangle$ is the change in SST averaged over the global ice-free ocean due to the $p$'th SST patch perturbation.  Because the patch covers only a small fraction of the ocean (especially if it is partly masked by land), this is much smaller than the SST perturbation at the patch center, $A_p = \pm 2$~K. For fully ocean-covered patches in the tropics, $\langle \Delta SST_p \rangle \approx 0.008 A_p$. We exclude patches where valid grid points, as described in Section \ref{sec:gfmip-simulations}, comprise less than 3\% of the patch. In these cases, the $\langle \Delta SST_p \rangle$ is less than 10$^{-5}$~K and the SST sensitivity cannot be reliably calculated.

Figure \ref{fig:results-ace-e3sm-sens-map} shows ACE and EAMv3's predicted radiative sensitivity $(d\overline{N}/dSST)_p$, constructed using warming and cooling patches separately (differenced from the control simulation), and using the difference of corresponding warm-patch and cool-patch simulations (which is also the average of the one-sided warm and cold-patch results).  Were the radiative response linear in the patch SST perturbation amplitude, and were internal variability a negligible effect on the patch estimates $\Delta \overline{N}_p$, these three maps would look identical for EAMv3, and similarly for ACE.  Comparing the maps across each row of Figure \ref{fig:results-ace-e3sm-sens-map}, we see that the linearity assumption approximately holds for both ACE and EAMv3 for tropical SST patches but is less accurate for extratropical SST patches. In contrast, BJ24's Figure 2 indicates that other climate models show the radiative sensitivity derived from the warm-patch simulations is generally more negative than that derived from the cool-patch simulations.

If ACE were a perfect emulator of EAMv3 and internal variability were negligible, all three map types would look the same for ACE as for EAMv3.  Comparing across each column of Fig. \ref{fig:results-ace-e3sm-sens-map}, we see qualitative similarity between ACE and EAMv3 in the radiative response maps for tropical SST perturbations but less agreement for midlatitude SST perturbations.  Like EAMv3, ACE produces a negative response to SST perturbations over the Indo-Pacific warm pool and tropical Atlantic, and a positive response to SST perturbations over the cooler subtropical ocean regions, consistent with physical understanding \cite{zhou_analyzing_2017}.  However, many details of the spatial structure of the SST sensitivity of global radiation are different between ACE and EAMv3.  For instance,  SST anomalies in and just north of the east Pacific ITCZ induce a positive global radiative response according to ACE, but not according to EAMv3.

Of these three map types, we prefer the 'average' maps based on differencing warm and cold patch simulation results for comparing emulator vs. model predictions in the present climate, for the following two reasons.
First, a centered finite difference is more accurate than a one-sided finite difference for estimating sensitivity to small perturbations, and the forced radiation `signal' $\Delta \overline{N}$ is twice as large, so it better rises about the `noise' associated with imperfect removal of internal climate variability by time averaging over a finite-length simulation, as discussed in the following section.  Second, all one-sided maps are constructed from differences of many 10-year patch simulations with the same 20-year control run.  Biases in the time-mean radiation fields in that control run will propagate to all the different fields and can have an undesirable systematic effect on the one-sided radiative response maps.

\subsection{Radiative response uncertainty due to internal variability}
\label{sec:uncertainty}
The GFMIP protocol bases the radiative response to the patch SST perturbations on time averages from many 10-year patch simulations, which are computationally affordable but contain residual impacts from internal variability.  Section 3.2.4 of BJ24 discusses that issue in support of their choices of patch and control simulation lengths, based on the assumptions (which we earlier verified for ACE and EAMv3) that TOA global-mean net radiation has comparable interannual variability for all patches and the control run, and that variability is uncorrelated from year to year. Their Equation 7 provides an uncertainty estimate for a one-sided (patch minus control) estimate of the radiative sensitivity $(d\overline{N}/dSST)_p$, reproduced here for convenience:
\begin{equation}
  \sigma_p^{1-sided} = \sigma_{\overline{N}}\left( \frac{1}{y_p} + \frac{1}{y_c} \right)^{1/2}/ |\langle \Delta SST_p \rangle|
\end{equation}
Here, $y_p = 10$ and $y_c = 20$ are the number of years of the GFMIP-specified patch and control simulations, respectively.  For ACE and EAMv3, we earlier estimated the interannual standard deviation of the patch and control runs to be $\sigma_{\overline{N}} = 0.2 W/m^2$, and the SST perturbation averaged over the entire ice-free ocean to be  $\langle \Delta SST_p \rangle = 0.008A_p$, where $A_p$ is the SST perturbation at the patch center ($\pm 2$~K for warm and cold patches, respectively). 

Based on these numbers, the estimated 1-sided uncertainty in warm and cold patch estimates of the radiative sensitivity
for ACE and EAMv3 is:
\begin{equation}
  \sigma_p^{ACE,1-sided} = \sigma_p^{EAMv3,1-sided} 
  \approx 5\, {\rm W}/{\rm m}^2/{\rm K}
  \label{eq:sigma_p_1sided_EAMv3}
\end{equation}
For patches that are partly masked by land, $\langle \Delta SST_p \rangle$ is smaller and the uncertainty correspondingly larger.

\begin{figure}[h]
    \centering
    \includegraphics[width=\textwidth]{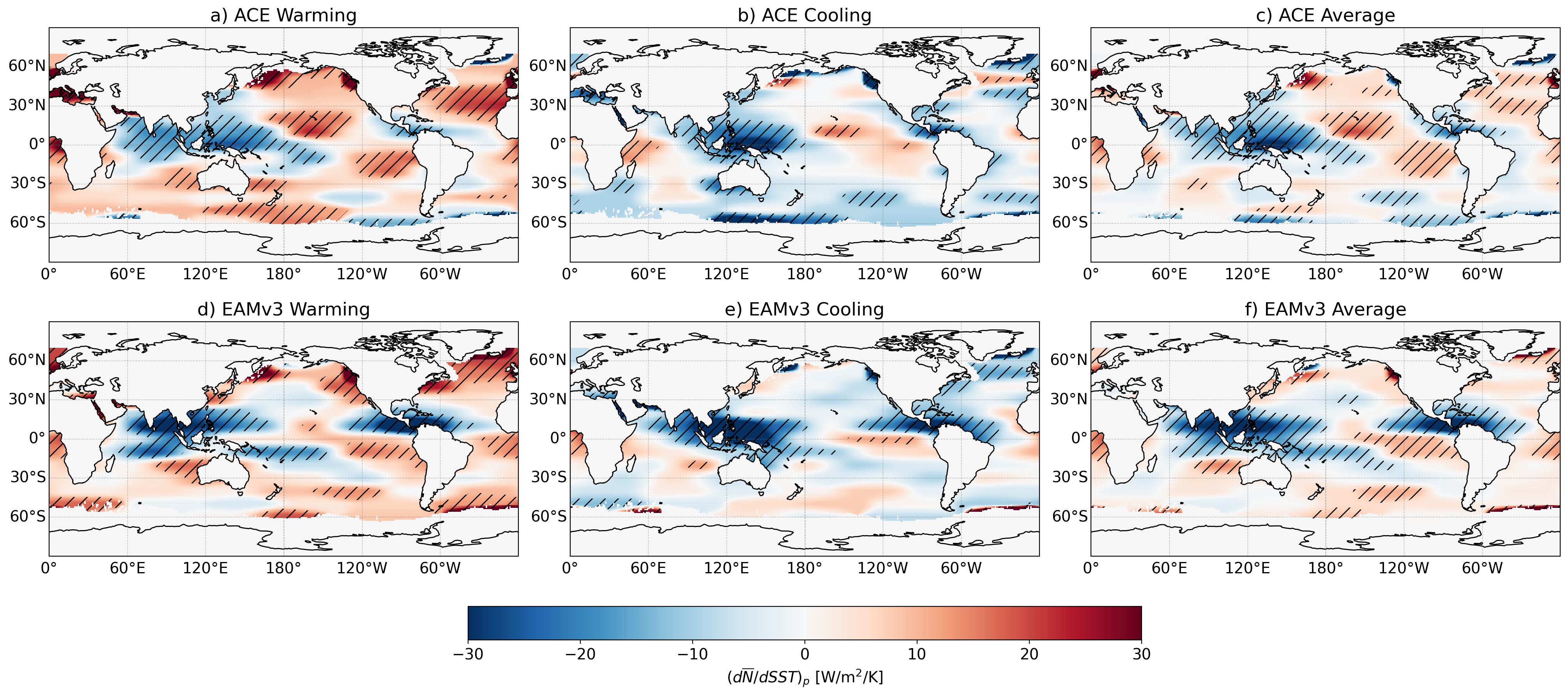}
    \caption{Normalized derivatives of TOA radiation with respect to change in SST using Green's function method. Warming denotes the estimate using only +2K patches, cooling uses -2K patches, and the average is the average of warming and cooling. Top row shows the estimate from ACE, while the bottom row is from EAMv3. Hatches indicate radiative sensitivities with magnitudes above approximate 95\% significance thresholds of 10~W/m$^2$ for panels a-b and d-e, and $6$~W/m$^2$ for panels c and f, derived in Sec. \ref {sec:uncertainty}.}
    \label{fig:results-ace-e3sm-sens-map}
\end{figure}

A similar analysis of the uncertainty of the 2-sided `average' estimate of radiative sensitivity made by differencing the warm and cold patch results gives
\begin{equation}
  \sigma_p^{av} = \sigma_{\overline{N}}\left( \frac{2}{y_p}\right)^{1/2}/ |2\langle \Delta SST_p \rangle|
\end{equation}
and the numerical estimate
\begin{equation}
  \sigma_p^{ACE,av} = \sigma_p^{EAMv3,av} 
  \approx 3\, {\rm W}/{\rm m}^2/{\rm K}
  \label{eq:sigma_p_av_EAMv3}
\end{equation}
In Fig. \ref{fig:results-ace-e3sm-sens-map}, darkened regions indicate radiative sensitivities with magnitude exceeding $2 \sigma_p$ based on the appropriate estimate of $\sigma_p$ (1-sided or average).  This is an approximate threshold for 95\% confidence that the radiative sensitivity is nonzero.  While the strongest signals far exceed this threshold, the sensitivities to SST perturbations over much of the extratropical oceans do not.  That is, we should not over-interpret spatial details of the radiative response maps generated by the GFMIP protocol.  

The central question for this paper is how well the radiative sensitivity map of ACE matches that of the EAMv3 model that it is emulating. Since the interannual variability should be uncorrelated between these models, we estimate the noise floor for their difference map (based on the average method) to be
\begin{equation}
  \sigma_p^{diff,av} = 2^{1/2}\sigma_p^{EAMv3,av} 
  \approx 4\,{\rm W}/{\rm m}^2/{\rm K}
  \label{eq:sigma_p_av_diff}
\end{equation}
This noise floor could be halved if we used 40 year (rather than 10 year) patch simulations, which is computationally straightforward for ACE but less so for EAMv3, since it requires almost 8000 simulation years.

\begin{figure}[h]
    \centering
    \includegraphics[width=0.5\textwidth]{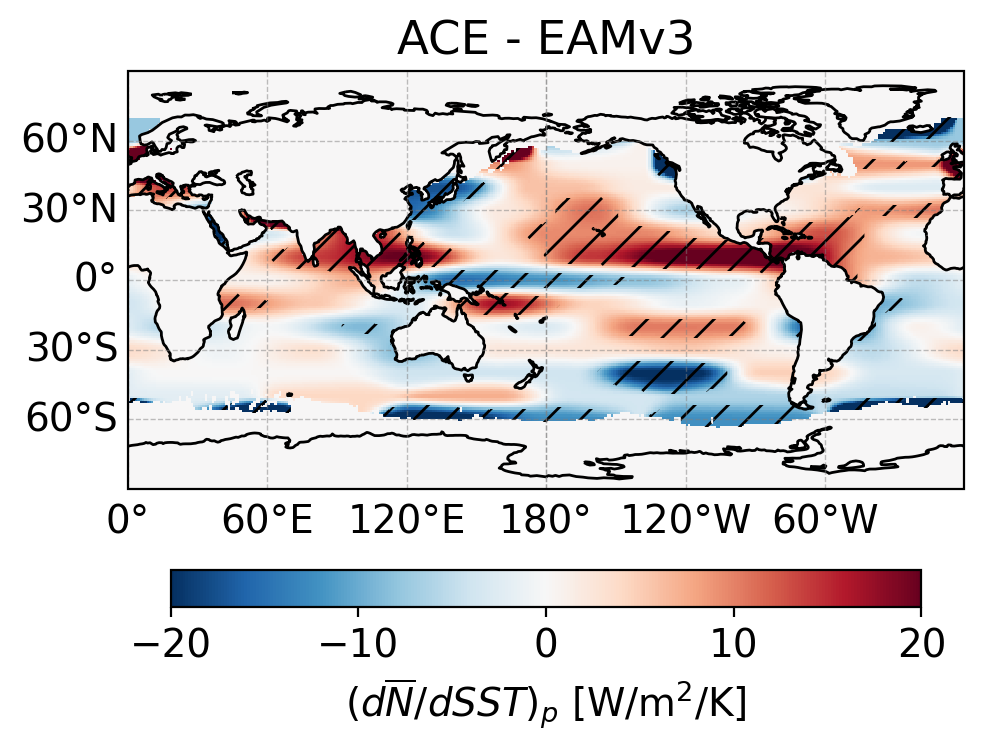}
    \caption{Difference between ACE and EAMv3 Green's function estimates of global TOA radiation response with SST, constructed by differencing 10-year warm and cold patch simulations. Hatches indicates SST patches for which this difference is greater than an approximate 95\% significance threshold of 8~W/m$^2$ derived in Sec. \ref{sec:uncertainty}.}
    \label{fig:results-ace-minus-e3sm-sens}
\end{figure}

Figure \ref{fig:results-ace-minus-e3sm-sens} shows the map of radiative sensitivity difference of ACE vs. EAMv3, where darkened regions indicate statistically significant magnitudes above $2\sigma_p^{diff,av}$. By this measure, ACE gives a biased radiative response vs. EAMv3 for many patches in the low-latitude oceans, with the biggest biases along the latitude band 0-20$^\circ$N in the tropical eastern Pacific ocean. Differences over parts of western Indo-Pacific region are also above the noise, though the magnitudes are smaller. Patches in southern Pacific ocean specifically around 40-50$^\circ$S are also biased. Despite these biases, the ACE and EAMv3 radiative sensitivity maps have a respectably positive area-weighted spatial pattern correlation of 0.53.

We hypothesize that the 51-year EAMv3 historical AMIP training data simulation may not sample tropical SST variability well enough to generalize to the range of SST perturbations used in GFMIP.  Augmenting the training to also sample from a long pre-industrial control run of EAMv3 might expand the range of SST variability that ACE sees in training and thereby reduce the apparent systematic emulator bias seen in Figure \ref{fig:results-ace-minus-e3sm-sens}.

\subsection{Historical reconstruction of TOA radiation}

SST Green's functions derived from the GFMIP protocol have caveats (e.g. linearity, internal variability, patch shape and size) that must be kept in mind when they are applied to the pattern effect. A consistency check that helps confirm their credibility is the accuracy of a Green's function reconstruction of the time series of historical annual-mean global net radiation, derived from the same climate model as was used to generate the Green's functions \cite{zhou_analyzing_2017,bloch-johnson_greens_2024}.  In this section, we apply this consistency check to ACE and EAMv3.

We superpose the patch-based Green's functions to estimate the change in $\overline{N}$ associated with an arbitrary historical anomaly $SST'(\phi,\theta,t)$ from some appropriate climatological mean.  We specialize Eqn. (5) of BJ24 to a grid comprised of the patch centers, which greatly simplifies the mathematics without changing the result. The patches can be used as a finite volume basis that reconstructs the $SST'$ field:
\begin{equation}
  SST'(\phi,\theta,t) = \sum_{p} SST'_p(t) \Delta SST_p(\phi,\theta,t)/A_p
  \label{eq:SSThistGF}
\end{equation}
where $p$ is the patch index and $SST'_p(t)$ is the SST anomaly at the center of patch $p$ (since the GFMIP specification ensures the patch center does not lie in the SST footprint of any other patches).  The factor $A_p$ accounts for the assumed patch-center amplitude of the SST patches.

For each patch we use the 'average' method to calculate the net global radiation difference between the warm and cold patch simulations, and divide it by two to get $\Delta \overline{N}^{avg}_p$.  We reconstruct the historical time series of net radiation using the same patch superposition: 
\begin{equation}
  \overline{N}'(t) \approx \sum_{p} SST'_p(t) \Delta \overline{N}^{avg}_p/A_p
  \label{eq:NhistGF}
\end{equation}

\begin{figure}[h]
    \centering
    \includegraphics[width=0.7\textwidth]{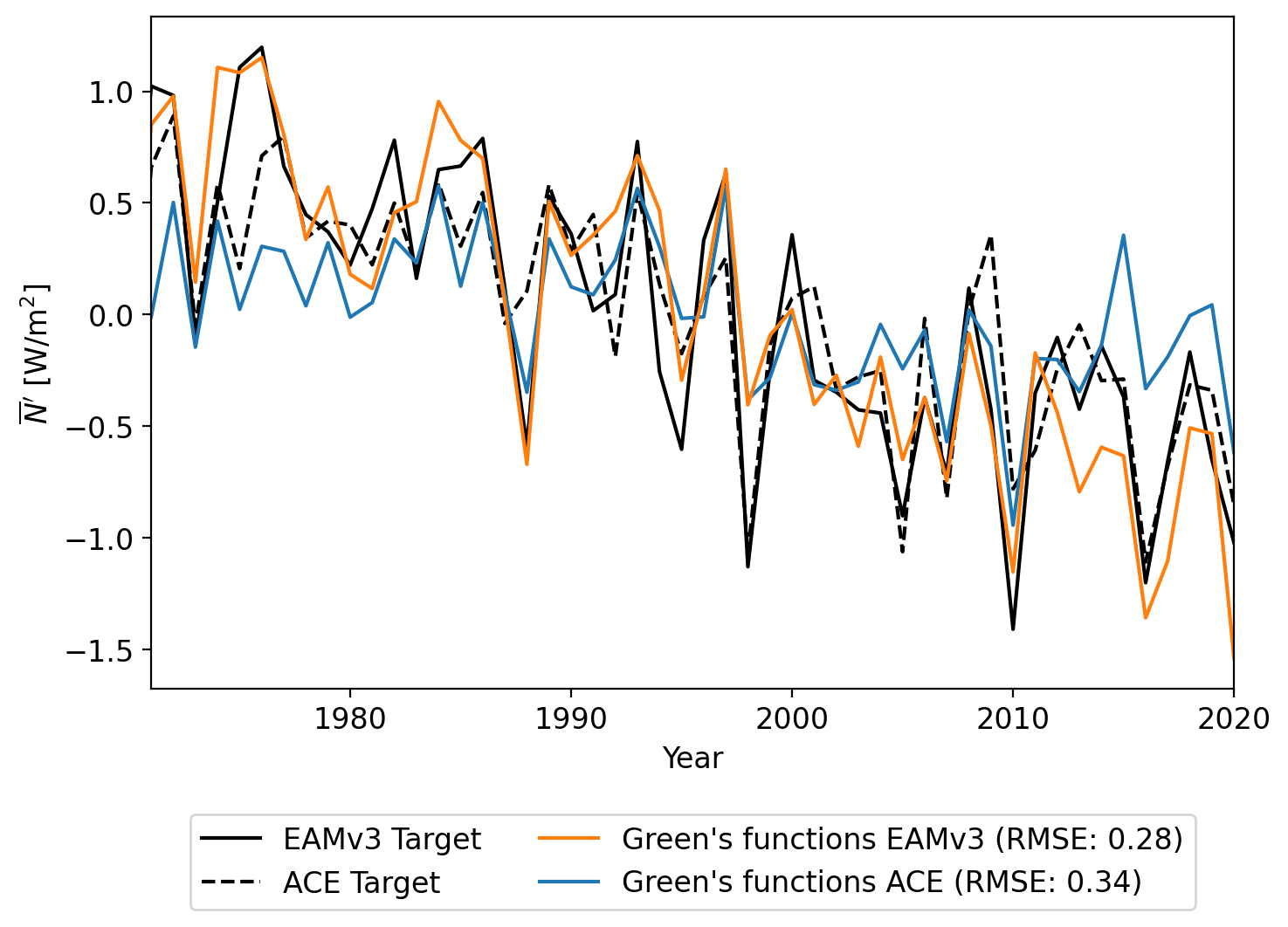}
    \caption{Historical reconstruction of TOA radiation estimated from Green's functions calculated from ACE (40 year patch simulations) and EAMv3 (10 year patch simulations) and multiplied by the sea surface temperature anomalies during 1970-2020 AMIP time period. The actual historical TOA radiation anomalies are plotted in solid and dash dark lines as EAMv3 and ACE target. RMSEs for the EAMv3 and ACE reconstructions are calculated vs. their respective historical targets. The RMSE for Green's functions ACE with respect to EAMv3 historical target is 0.43. \label{fig:results-hist-reconstruc-dR}}
\end{figure}

To evaluate how well the patch Green's functions derived from ACE and EAMv3 reconstruct their respective historical TOA radiation time series, we consider the training data from section \ref{sec:ace-eamv3-training} as the historical simulation. We consider $\overline{N}'(t)$ as the anomalies with respect to the time mean of the full EAMv3 simulation from 1970-2020, plotted as the black line in Figure \ref{fig:results-hist-reconstruc-dR}. We also show $\overline{N}'(t)$ from ACE's historical rollout for the same time period, plotted as the dashed black line. We then reconstruct $\Delta \overline{N}$ following Equation \ref{eq:NhistGF} 
choosing $SST'(\phi,\theta,t)$ to be the sea surface temperature anomalies from the historical average over 1970-2020.  The EAMv3 patch simulations are for 10 years.  However, since we are trying to isolate an SST-forced signal, and ACE is computationally efficient, we use 40 year ACE patch simulations to reduce the impact of internal variability.

Figure \ref{fig:results-hist-reconstruc-dR} shows that both ACE and EAMv3 Green's function reconstructions capture the historical $\overline{N}'(t)$ remarkably well when compared to its respective target. Both ACE and EAMv3 show a decreasing trend of $\Delta \overline{N}$, which is expected since the mean global temperature increases during this period resulting in an increase in outgoing energy flux. The Green's function reconstruction from ACE has a smaller interannual standard deviation ($\sigma$=0.32 W/m$^2$) compared to its target ($\sigma$=0.50 W/m$^2$), while the reconstruction from EAMv3 ($\sigma$=0.66 W/m$^2$) matches its target ($\sigma$=0.63 W/m$^2$) much closer.

\section{Conclusions}

We carried out a Green's function experiment following the GFMIP protocol of BJ24, comparing the EAMv3 physics-based global atmosphere model with an ACE emulator of this model that runs 100 times faster.  Our goal was to investigate whether an emulator is ready for this task. We first trained ACE using a 1970-2020 AMIP-style simulation of EAMv3 (ACE-EAMv3) then used this version of ACE to autoregressively (with 6 hour rollout steps) generate patch simulations of +/-2K sea surface temperature perturbations. We also performed the same patch simulations using EAMv3. Since we could run ACE cheaply, we extended the patch simulations to 40 years; a longer time average reduces the 'noise' effects of interannual variability. We found that ACE simulates the spatial pattern of an individual patch's TOA radiative flux response well.  However, the global mean radiative flux response to a given SST patch involves strong cancellation between regions of positive and negative radiative flux responses, so it is much more challenging for the emulator to capture to high relative accuracy.

We constructed Green's function sensitivity maps of TOA global net radiation to SST perturbations for all patches for ACE and EAMv3.
They were qualitatively similar in many places, but there were noticeable discrepancies in spatial details that significantly exceed estimated `noise thresholds' due to uncertainties in 10-year means given natural internal variability. The largest discrepancy between ACE and EAMv3 occurs over the northeast tropical Pacific. A potential contributor to the discrepancies is that the AMIP reference simulation used to train ACE does not have a sufficiently diverse set of interannual SST anomaly patterns to tightly constrain all the SST patch responses.  This could be tested by training ACE directly on a full suite of patch simulations to see if this improves the skill of its radiative sensitivity maps.  Unfortunately, that would require large volumes of EAMv3 model output that were not saved in conducting this study. 

Despite these biases, we find Green's function reconstructions of the 1970-2020 global annual-mean TOA radiative flux using ACE and EAMv3 approximately reproduce the trend and variability in their respective AMIP simulations. This is an important improvement over the results with earlier versions of ACE reported by \citeA{loon_reanalysis-based_2025}, although we would not yet recommend ACE with our present AMIP training protocol as a substitute for deriving the Green's functions from a physical climate model. We are optimistic that ACE's remaining biases for this challenging but attractive application can be overcome in the near future.

\section*{Open Research Section}
The code used for model training and evaluation is available at \url{https://github.com/ai2cm/ace}. The scripts used for submitting experiments and generating figures are available at \url{https://github.com/ai2cm/ace-gfmip-paper}. The checkpoint used in this manuscript can be found on Hugging Face along with sample reference forcing data at \url{https://huggingface.co/allenai/ACE2-EAMv3}.

\acknowledgments
Lawrence Livermore National Laboratory authors were supported by Laboratory Directed Research and Development (LDRD 22-ERD-052), and Ai2 authors were supported by a subcontract included in this funding, as well as general funding for Ai2 from the Paul G. Allen estate.  This research used resources of the National Energy Research Scientific Computing Center (NERSC), a Department of Energy Office of Science User Facility using NERSC award BER-ERCAP0026743. C.P. and P.P. were supported by Department of Energy (DOE) Award DE-SC0022110, through the Regional Modeling and Analysis Program (RGMA). J.N. and D.M. were supported by the U.S. Department of Energy, Office of Science, Office of Biological and Environmental Research, through the Established Program to Stimulate Competitive Research (DOE EPSCoR) under Award Number DE-SC0024161. We would like to acknowledge the use of computational resources (doi:10.5065/D6RX99HX) at the NCAR-Wyoming Supercomputing Center provided by the National Science Foundation and the State of Wyoming, and supported by NCAR's Computational and Information Systems Laboratory  through the Wyoming-NCAR alliance.

\bibliography{main}

\begin{thebibliography}{}

\bibitem [\protect \citeauthoryear {%
Alessi%
\ \BBA {} Rugenstein%
}{%
Alessi%
\ \BBA {} Rugenstein%
}{%
{\protect \APACyear {2023}}%
}]{%
alessi_surface_2023}
\APACinsertmetastar {%
alessi_surface_2023}%
\begin{APACrefauthors}%
Alessi, M\BPBI J.%
\BCBT {}\ \BBA {} Rugenstein, M\BPBI A\BPBI A.%
\end{APACrefauthors}%
\unskip\
\newblock
\APACrefYearMonthDay{2023}{}{}.
\newblock
{\BBOQ}\APACrefatitle {Surface {Temperature} {Pattern} {Scenarios} {Suggest} {Higher} {Warming} {Rates} {Than} {Current} {Projections}} {Surface {Temperature} {Pattern} {Scenarios} {Suggest} {Higher} {Warming} {Rates} {Than} {Current} {Projections}}.{\BBCQ}
\newblock
\APACjournalVolNumPages{Geophysical Research Letters}{50}{23}{e2023GL105795}.
\newblock
\begin{APACrefDOI} \doi{10.1029/2023GL105795} \end{APACrefDOI}
\PrintBackRefs{\CurrentBib}

\bibitem [\protect \citeauthoryear {%
Barsugli%
\ \BBA {} Sardeshmukh%
}{%
Barsugli%
\ \BBA {} Sardeshmukh%
}{%
{\protect \APACyear {2002}}%
}]{%
barsugli_global_2002}
\APACinsertmetastar {%
barsugli_global_2002}%
\begin{APACrefauthors}%
Barsugli, J\BPBI J.%
\BCBT {}\ \BBA {} Sardeshmukh, P\BPBI D.%
\end{APACrefauthors}%
\unskip\
\newblock
\APACrefYearMonthDay{2002}{}{}.
\newblock
{\BBOQ}\APACrefatitle {Global {Atmospheric} {Sensitivity} to {Tropical} {SST} {Anomalies} throughout the {Indo}-{Pacific} {Basin}} {Global {Atmospheric} {Sensitivity} to {Tropical} {SST} {Anomalies} throughout the {Indo}-{Pacific} {Basin}}.{\BBCQ}
\newblock
\APACjournalVolNumPages{Journal of Climate}{}{}{}.
\PrintBackRefs{\CurrentBib}

\bibitem [\protect \citeauthoryear {%
Bloch-Johnson%
\ \protect \BOthers {.}}{%
Bloch-Johnson%
\ \protect \BOthers {.}}{%
{\protect \APACyear {2024}}%
}]{%
bloch-johnson_greens_2024}
\APACinsertmetastar {%
bloch-johnson_greens_2024}%
\begin{APACrefauthors}%
Bloch-Johnson, J.%
, Rugenstein, M\BPBI A\BPBI A.%
, Alessi, M\BPBI J.%
, Proistosescu, C.%
, Zhao, M.%
, Zhang, B.%
\BDBL {}Zhou, C.%
\end{APACrefauthors}%
\unskip\
\newblock
\APACrefYearMonthDay{2024}{}{}.
\newblock
{\BBOQ}\APACrefatitle {The {Green}'s {Function} {Model} {Intercomparison} {Project} ({GFMIP}) {Protocol}} {The {Green}'s {Function} {Model} {Intercomparison} {Project} ({GFMIP}) {Protocol}}.{\BBCQ}
\newblock
\APACjournalVolNumPages{Journal of Advances in Modeling Earth Systems}{16}{2}{e2023MS003700}.
\newblock
\begin{APACrefDOI} \doi{10.1029/2023MS003700} \end{APACrefDOI}
\PrintBackRefs{\CurrentBib}

\bibitem [\protect \citeauthoryear {%
Branstator%
}{%
Branstator%
}{%
{\protect \APACyear {1985}}%
}]{%
branstator_analysis_1985}
\APACinsertmetastar {%
branstator_analysis_1985}%
\begin{APACrefauthors}%
Branstator, G.%
\end{APACrefauthors}%
\unskip\
\newblock
\APACrefYearMonthDay{1985}{}{}.
\newblock
{\BBOQ}\APACrefatitle {Analysis of {General} {Circulation} {Model} {Sea}-{Surface} {Temperature} {Anomaly} {Simulations} {Using} a {Linear} {Model}. {Part} {I}: {Forced} {Solutions}} {Analysis of {General} {Circulation} {Model} {Sea}-{Surface} {Temperature} {Anomaly} {Simulations} {Using} a {Linear} {Model}. {Part} {I}: {Forced} {Solutions}}.{\BBCQ}
\newblock
\APACjournalVolNumPages{Journal of the Atmospheric Sciences}{}{}{}.
\PrintBackRefs{\CurrentBib}

\bibitem [\protect \citeauthoryear {%
Clark%
\ \protect \BOthers {.}}{%
Clark%
\ \protect \BOthers {.}}{%
{\protect \APACyear {2024}}%
}]{%
clark_ace2-som_2024}
\APACinsertmetastar {%
clark_ace2-som_2024}%
\begin{APACrefauthors}%
Clark, S\BPBI K.%
, Watt-Meyer, O.%
, Kwa, A.%
, McGibbon, J.%
, Henn, B.%
, Perkins, W\BPBI A.%
\BDBL {}Bretherton, C\BPBI S.%
\end{APACrefauthors}%
\unskip\
\newblock
\APACrefYearMonthDay{2024}{}{}.
\newblock
\APACrefbtitle {{ACE2}-{SOM}: {Coupling} an {ML} atmospheric emulator to a slab ocean and learning the sensitivity of climate to changed CO2.} {{ACE2}-{SOM}: {Coupling} an {ML} atmospheric emulator to a slab ocean and learning the sensitivity of climate to changed co2.}
\newblock
\APACaddressPublisher{}{arXiv}.
\newblock
\APACrefnote{arXiv:2412.04418 [physics]}
\newblock
\begin{APACrefDOI} \doi{10.48550/arXiv.2412.04418} \end{APACrefDOI}
\PrintBackRefs{\CurrentBib}

\bibitem [\protect \citeauthoryear {%
Dong%
, Proistosescu%
, Armour%
\BCBL {}\ \BBA {} Battisti%
}{%
Dong%
\ \protect \BOthers {.}}{%
{\protect \APACyear {2019}}%
}]{%
dong_attributing_2019}
\APACinsertmetastar {%
dong_attributing_2019}%
\begin{APACrefauthors}%
Dong, Y.%
, Proistosescu, C.%
, Armour, K\BPBI C.%
\BCBL {}\ \BBA {} Battisti, D\BPBI S.%
\end{APACrefauthors}%
\unskip\
\newblock
\APACrefYearMonthDay{2019}{}{}.
\newblock
{\BBOQ}\APACrefatitle {Attributing {Historical} and {Future} {Evolution} of {Radiative} {Feedbacks} to {Regional} {Warming} {Patterns} using a {Green}’s {Function} {Approach}: {The} {Preeminence} of the {Western} {Pacific}} {Attributing {Historical} and {Future} {Evolution} of {Radiative} {Feedbacks} to {Regional} {Warming} {Patterns} using a {Green}’s {Function} {Approach}: {The} {Preeminence} of the {Western} {Pacific}}.{\BBCQ}
\newblock
\APACjournalVolNumPages{Journal of Climate}{}{}{}.
\newblock
\begin{APACrefDOI} \doi{10.1175/JCLI-D-18-0843.1} \end{APACrefDOI}
\PrintBackRefs{\CurrentBib}

\bibitem [\protect \citeauthoryear {%
Duncan%
\ \protect \BOthers {.}}{%
Duncan%
\ \protect \BOthers {.}}{%
{\protect \APACyear {2024}}%
}]{%
Duncan2024}
\APACinsertmetastar {%
Duncan2024}%
\begin{APACrefauthors}%
Duncan, J\BPBI P\BPBI C.%
, Wu, E.%
, Golaz, J.%
, Caldwell, P\BPBI M.%
, Watt‐Meyer, O.%
, Clark, S\BPBI K.%
\BDBL {}Bretherton, C\BPBI S.%
\end{APACrefauthors}%
\unskip\
\newblock
\APACrefYearMonthDay{2024}{}{}.
\newblock
{\BBOQ}\APACrefatitle {Application of the AI2 Climate Emulator to E3SMv2’s Global Atmosphere Model, With a Focus on Precipitation Fidelity} {Application of the ai2 climate emulator to e3smv2’s global atmosphere model, with a focus on precipitation fidelity}.{\BBCQ}
\newblock
\APACjournalVolNumPages{Journal of Geophysical Research: Machine Learning and Computation}{1}{3}{}.
\newblock
\begin{APACrefDOI} \doi{10.1029/2024jh000136} \end{APACrefDOI}
\PrintBackRefs{\CurrentBib}

\bibitem [\protect \citeauthoryear {%
Eyring%
\ \protect \BOthers {.}}{%
Eyring%
\ \protect \BOthers {.}}{%
{\protect \APACyear {2016}}%
}]{%
Eyring2016}
\APACinsertmetastar {%
Eyring2016}%
\begin{APACrefauthors}%
Eyring, V.%
, Bony, S.%
, Meehl, G\BPBI A.%
, Senior, C\BPBI A.%
, Stevens, B.%
, Stouffer, R\BPBI J.%
\BCBL {}\ \BBA {} Taylor, K\BPBI E.%
\end{APACrefauthors}%
\unskip\
\newblock
\APACrefYearMonthDay{2016}{}{}.
\newblock
{\BBOQ}\APACrefatitle {Overview of the Coupled Model Intercomparison Project Phase 6 (CMIP6) experimental design and organization} {Overview of the coupled model intercomparison project phase 6 (cmip6) experimental design and organization}.{\BBCQ}
\newblock
\APACjournalVolNumPages{Geoscientific Model Development}{9}{5}{1937–1958}.
\newblock
\begin{APACrefDOI} \doi{10.5194/gmd-9-1937-2016} \end{APACrefDOI}
\PrintBackRefs{\CurrentBib}

\bibitem [\protect \citeauthoryear {%
Gates%
\ \protect \BOthers {.}}{%
Gates%
\ \protect \BOthers {.}}{%
{\protect \APACyear {1999}}%
}]{%
gates_overview_1999}
\APACinsertmetastar {%
gates_overview_1999}%
\begin{APACrefauthors}%
Gates, W\BPBI L.%
, Boyle, J\BPBI S.%
, Covey, C.%
, Dease, C\BPBI G.%
, Doutriaux, C\BPBI M.%
, Drach, R\BPBI S.%
\BDBL {}Williams, D\BPBI N.%
\end{APACrefauthors}%
\unskip\
\newblock
\APACrefYearMonthDay{1999}{}{}.
\newblock
{\BBOQ}\APACrefatitle {An {Overview} of the {Results} of the {Atmospheric} {Model} {Intercomparison} {Project} ({AMIP} {I})} {An {Overview} of the {Results} of the {Atmospheric} {Model} {Intercomparison} {Project} ({AMIP} {I})}.{\BBCQ}
\newblock
\APACjournalVolNumPages{Bulletin of the American Meteorological Society}{}{}{}.
\PrintBackRefs{\CurrentBib}

\bibitem [\protect \citeauthoryear {%
Loon%
, Rugenstein%
\BCBL {}\ \BBA {} Barnes%
}{%
Loon%
\ \protect \BOthers {.}}{%
{\protect \APACyear {2025}}%
}]{%
loon_reanalysis-based_2025}
\APACinsertmetastar {%
loon_reanalysis-based_2025}%
\begin{APACrefauthors}%
Loon, S\BPBI V.%
, Rugenstein, M.%
\BCBL {}\ \BBA {} Barnes, E\BPBI A.%
\end{APACrefauthors}%
\unskip\
\newblock
\APACrefYearMonthDay{2025}{}{}.
\newblock
\APACrefbtitle {Reanalysis-based {Global} {Radiative} {Response} to {Sea} {Surface} {Temperature} {Patterns}: {Evaluating} the {Ai2} {Climate} {Emulator}.} {Reanalysis-based {Global} {Radiative} {Response} to {Sea} {Surface} {Temperature} {Patterns}: {Evaluating} the {Ai2} {Climate} {Emulator}.}
\newblock
\APACaddressPublisher{}{arXiv}.
\newblock
\begin{APACrefDOI} \doi{10.48550/arXiv.2502.10893} \end{APACrefDOI}
\PrintBackRefs{\CurrentBib}

\bibitem [\protect \citeauthoryear {%
Watt-Meyer%
\ \protect \BOthers {.}}{%
Watt-Meyer%
\ \protect \BOthers {.}}{%
{\protect \APACyear {2023}}%
}]{%
WattMeyer2023}
\APACinsertmetastar {%
WattMeyer2023}%
\begin{APACrefauthors}%
Watt-Meyer, O.%
, Dresdner, G.%
, McGibbon, J.%
, Clark, S\BPBI K.%
, Henn, B.%
, Duncan, J.%
\BDBL {}Bretherton, C\BPBI S.%
\end{APACrefauthors}%
\unskip\
\newblock
\APACrefYearMonthDay{2023}{}{}.
\newblock
\APACrefbtitle {{ACE: A} fast, skillful learned global atmospheric model for climate prediction.} {{ACE: A} fast, skillful learned global atmospheric model for climate prediction.}
\newblock
\APACaddressPublisher{}{arXiv}.
\newblock
\begin{APACrefDOI} \doi{10.48550/arxiv.2310.02074} \end{APACrefDOI}
\PrintBackRefs{\CurrentBib}

\bibitem [\protect \citeauthoryear {%
Watt-Meyer%
\ \protect \BOthers {.}}{%
Watt-Meyer%
\ \protect \BOthers {.}}{%
{\protect \APACyear {2024}}%
}]{%
watt-meyer_ace2_2024}
\APACinsertmetastar {%
watt-meyer_ace2_2024}%
\begin{APACrefauthors}%
Watt-Meyer, O.%
, Henn, B.%
, McGibbon, J.%
, Clark, S\BPBI K.%
, Kwa, A.%
, Perkins, W\BPBI A.%
\BDBL {}Bretherton, C\BPBI S.%
\end{APACrefauthors}%
\unskip\
\newblock
\APACrefYearMonthDay{2024}{}{}.
\newblock
\APACrefbtitle {{ACE2}: {Accurately} learning subseasonal to decadal atmospheric variability and forced responses.} {{ACE2}: {Accurately} learning subseasonal to decadal atmospheric variability and forced responses.}
\newblock
\APACaddressPublisher{}{arXiv}.
\newblock
\begin{APACrefDOI} \doi{10.48550/arXiv.2411.11268} \end{APACrefDOI}
\PrintBackRefs{\CurrentBib}

\bibitem [\protect \citeauthoryear {%
Xie%
\ \protect \BOthers {.}}{%
Xie%
\ \protect \BOthers {.}}{%
{\protect \APACyear {2025}}%
}]{%
xie2025energy}
\APACinsertmetastar {%
xie2025energy}%
\begin{APACrefauthors}%
Xie, S.%
, Terai, C\BPBI R.%
, Wang, H.%
, Tang, Q.%
, Fan, J.%
, Burrows, S\BPBI M.%
\BDBL {}others%
\end{APACrefauthors}%
\unskip\
\newblock
\APACrefYearMonthDay{2025}{}{}.
\newblock
{\BBOQ}\APACrefatitle {The Energy Exascale Earth System Model Version 3. Part I: Overview of the Atmospheric Component} {The energy exascale earth system model version 3. part i: Overview of the atmospheric component}.{\BBCQ}
\newblock
\APACjournalVolNumPages{Authorea Preprints}{}{}{}.
\newblock
\begin{APACrefDOI} \doi{10.22541/essoar.174456922.21825772/v1} \end{APACrefDOI}
\PrintBackRefs{\CurrentBib}

\bibitem [\protect \citeauthoryear {%
Zhang%
, Zhao%
\BCBL {}\ \BBA {} Tan%
}{%
Zhang%
\ \protect \BOthers {.}}{%
{\protect \APACyear {2023}}%
}]{%
zhang_using_2023}
\APACinsertmetastar {%
zhang_using_2023}%
\begin{APACrefauthors}%
Zhang, B.%
, Zhao, M.%
\BCBL {}\ \BBA {} Tan, Z.%
\end{APACrefauthors}%
\unskip\
\newblock
\APACrefYearMonthDay{2023}{}{}.
\newblock
{\BBOQ}\APACrefatitle {Using a {Green}’s {Function} {Approach} to {Diagnose} the {Pattern} {Effect} in {GFDL} {AM4} and {CM4}} {Using a {Green}’s {Function} {Approach} to {Diagnose} the {Pattern} {Effect} in {GFDL} {AM4} and {CM4}}.{\BBCQ}
\newblock
\APACjournalVolNumPages{Journal of Climate}{}{}{}.
\newblock
\begin{APACrefDOI} \doi{10.1175/JCLI-D-22-0024.1} \end{APACrefDOI}
\PrintBackRefs{\CurrentBib}

\bibitem [\protect \citeauthoryear {%
Zhou%
, Zelinka%
\BCBL {}\ \BBA {} Klein%
}{%
Zhou%
\ \protect \BOthers {.}}{%
{\protect \APACyear {2017}}%
}]{%
zhou_analyzing_2017}
\APACinsertmetastar {%
zhou_analyzing_2017}%
\begin{APACrefauthors}%
Zhou, C.%
, Zelinka, M\BPBI D.%
\BCBL {}\ \BBA {} Klein, S\BPBI A.%
\end{APACrefauthors}%
\unskip\
\newblock
\APACrefYearMonthDay{2017}{}{}.
\newblock
{\BBOQ}\APACrefatitle {Analyzing the dependence of global cloud feedback on the spatial pattern of sea surface temperature change with a {Green}'s function approach} {Analyzing the dependence of global cloud feedback on the spatial pattern of sea surface temperature change with a {Green}'s function approach}.{\BBCQ}
\newblock
\APACjournalVolNumPages{Journal of Advances in Modeling Earth Systems}{9}{5}{}.
\newblock
\begin{APACrefDOI} \doi{10.1002/2017MS001096} \end{APACrefDOI}
\PrintBackRefs{\CurrentBib}

\end{thebibliography}

\end{document}